# A Quantitative Understanding of Human Sex Chromosomal Genes

Sk. Sarif Hassan[1], Pabitra Pal Choudhury[2], Antara Sengupta[2], Binayak Sahu[*], Rojalin Mishra[*], Devendra Kumar Yadav[*], Saswatee Panda[*], Dharamveer Pradhan[*], Shrusti Dash[*] and Gourav Pradhan[*]

[1] *International Centre for Theoretical Sciences, Tata Institute of Fundamental Research, Bangalore 560012, India*

[2,] *Applied Statistics Unit, Indian Statistical Institute, Calcutta 700108, India*

* *Visiting Student at Institute of Mathematics & Applications, Bhubaneswar 751003, Odisha, India*

*Correspondence to Sk. Sarif Hassan (sarif.hassan@icts.res.in)*

## Abstract:

In the last few decades, the human allosomes are engrossed in an intensive attention among researchers. The allosomes are now already been sequenced and found there are about 2000 and 78 genes in human X and Y chromosomes respectively. The hemizygosity of the human X chromosome in males exposes recessive disease alleles, and this phenomenon has prompted decades of intensive study of X-linked disorders. By contrast, the small size of the human Y chromosome, and its prominent long-arm heterochromatic region suggested absence of function beyond sex determination. But the present problem is to accomplish whether a given sequence of nucleotides i.e. a DNA is a Human X or Y chromosomal genes or not, without any biological experimental support. In our perspective, a proper quantitative understanding of these genes is required to justify or nullify whether a given sequence is a Human X or Y chromosomal gene. In this paper, some of the X and Y chromosomal genes have been quantified in genomic and proteomic level through Fractal Geometric and Mathematical Morphometric analysis. Using the proposed quantitative model, one can easily make probable justification or deterministic nullification whether a given sequence of nucleotides is a probable Human X or Y chromosomal gene or not, without seeking any biological experiment. Of course, a further biological experiment is essential to validate it as the probable Human X or Y chromosomal gene homologue. This study would enable Biologists to understand these genes in more quantitative manner instead of their qualitative features.

## 1. Introduction

In the last ten years, Genomics has revolutionized the study of evolution. Evolution changes the sequence of DNA molecules, and comparing DNA sequences allow us to reconstruct evolutionary events from the past. The availability of DNA sequences from multiple vertebrates has confirmed that the process of sex chromosome evolution as envisioned by theorists has played out multiple times in the evolution of vertebrate sexchromosomes. However, complete, high-quality sequences of sex chromosomes have led to discoveries that were unanticipated by existing theory. The next stage of genomic research will begin to derive meaningful knowledge from these genes. A quantitative genomic understanding will have a major impact in the fields of medicine, biotechnology, and the life sciences [1 and 2].

One of the most frontier contemporary challenges is to make a revolution in medical science by introducing *Genetic Therapy* [3]. *Gene therapy* is an experimental technique that uses genes to treat or prevent disease. This method of therapy would countenance us to treat a disorder by inserting a gene into a patient's cells instead of using drugs or surgery. The most commonly practiced approaches of gene therapy include…

- Replacing a mutated gene that causes disease with a healthy copy of the gene.
- Inactivating, or "knocking out," a mutated gene that is functioning improperly.
- Introducing a new gene into the body to help fight a disease.

Although gene therapy is a promising treatment option for a number of diseases (including inherited disorders, some types of cancer, and certain viral infections), the technique remains risky and is still under study to make sure that it will be safe and effective. Gene therapy is currently being tested for the treatment of diseases that have no other cures [3, 4, 5 and 6]. Prior to gene therapy as a practical approach for treating diseases, we must overcome many technical challenges. One of the nontrivial challenges is to get quantitative insight of genes. This would help us in precise characterization of a particular DNA. This quantitative study of genes will be an add-on as the genetic signature of a gene.

In the present study, a mathematical quantification of human X and Y chromosomal genes [7, 8, 9, and 10, 12 and 13] has been done by using *Fractal Geometry* [14, 15 and 16]. So on using this proposed quantitative model, one can easily make probable justification (or deterministic nullification) whether a given sequence of nucleotides is a probable Human X/Y chromosomal genes or its homologue or not, without seeking any biological experiment. This study would help researchers in understanding these genes in differentiating from each other through the very nucleotide syntactical presentation.

### *1.1 Model Decomposition and Representation*

*(A) DNA 4-Colored Representation:* Let a DNA sequence is in the form of four-letter (ATGC) nucleotides sequence (*Fig. 1A*). Such sequence shown in *Fig. 1A* is converted as a function (*Fig. 1B*) depicting colors Red, Blue, Green, and Yellow respectively for A, T, G, and C [17, 18]. This allows $f(x,y)$ having maximum of 4 colors, i.e. $0 \leq f(x,y) \leq 3$.

```
GTTGAGGGGGTGTTGAGGGCGGAGAAATGCAAGTTTCATTACAAAAGTTAACGTAACAAAGAATCTGGTAGAAGT
GAGTTTTGGATAGTAAAATAAGTTTCGAACTCTGGCACCTTTCAATTTTGTCGCACTCTCCTTGTTTTTGACAAT
GCAATCATATGCTTCTGCTATGTTAAGCGTATTCAACAGCGATGATTACAGTCCAGCTGTGCAAGAGAATATTCC
CGCTCTCCGGAGAAGCTCTTCCTTCCTTTGCACTGAAAGCTGTAACTCTAAGTATCAGTGTGAAACGGGAGAAAA
CAGTAAAGGCAACGTCCAGGATAGAGTGAAGCGACCCATGAACGCATTCATCGTGTGGTCTCGCGATCAGAGGCG
CAAGATGGCTCTAGAGAATCCCAGAATGCGAAACTCAGAGATCAGCAAGCAGCTGGGATACCAGTGGAAAATGCT
TACTGAAGCCGAAAAATGGCCATTCTTCCAGGAGGCACAGAAATTACAGGCCATGCACAGAGAGAAATACCCGAA
TTATAAGTATCGACCTCGTCGGAAGGCGAAGATGCTGCCGAAGAATTGCAGTTTGCTTCCCGCAGATCCCGCTTC
GGTACTCTGCAGCGAAGTGCAACTGGACAACAGGTTGTACAGGGATGACTGTACGAAAGCCACACACTCAAGAAT
GGAGCACCAGCTAGGCCACTTACCGCCCATCAACGCAGCCAGCTCACCGCAGCAACGGGACCGCTACAGCCACTG
GACAAAGCTGTAGGACAATCGGGTAACATTGGCTACAAAGACCTACCTAGATGCTCCTTTTTACGATAACTTACA
GCCCTCACTTTCTTATGTTTAGTTTCAATATTGTTTTCTTTTCTCTGGCTAATAAAGGCCTTATTCATTTCA
```

*Fig. 1 (A): A DNA string of four variables A T C and G of SRY*

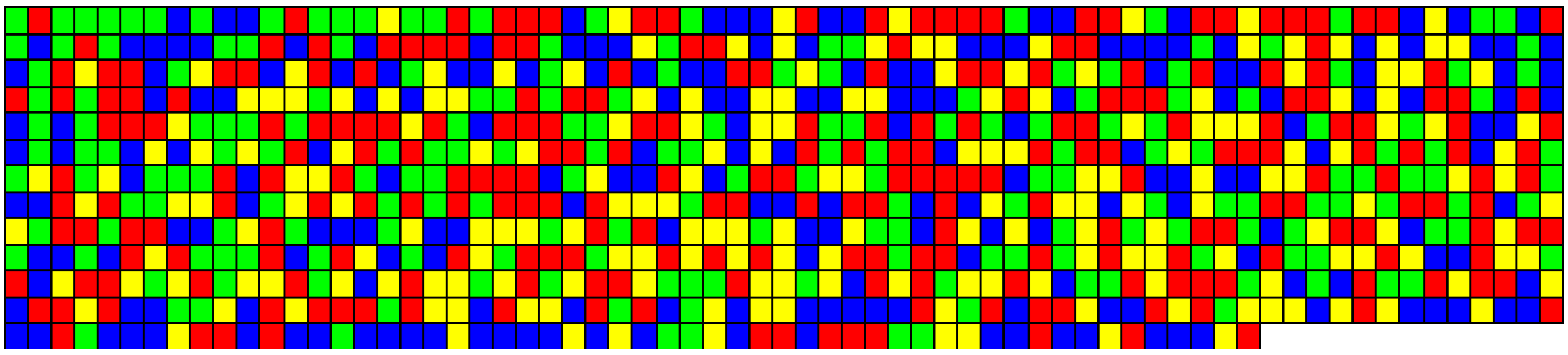

*Fig.1. (B) Function generated by proper colour coding for ATGC of SRY*

*(B) Binary Representation*: A DNA as a one dimensional nucleotide sequence, and is represented as a map such that $T(A) = 00$; $T(T) = 11$, $T(C) = 01$ *and* $T(G) = 10$. This mapping yields a DNA sequence in a binary string format. A portion of such a binary string is shown below of some fixed size (twice of the DNA sequence length). We call this representation of DNA as *2-adic string* of DNA [18].

00111000010010100011111000000000000011000010000011110001000100111100111101011111110000010011
11100010111111010101001001111111110000010110010011110000000011000011110011001101100100110000
00000001111111101111001100111111111110100011111110111110011111111101000000000011110011000000
00111110101011101100001000010011111011111010011011001…..(some more 0, 1 are there in the string).

*(C) 4-Adic Representation*: Also we consider a DNA as a string of four variables 0, 1, 2 and 3 (as shown below) corresponding to A T C and G respectively. We name this string as *4-adic string* of DNA [17, 18].

0230103302230000020030022010102202211222001022303222111031222300320312300002
002202021310200000122232202022222101222122022221000002202000022333232003010
221220022120030000232230222231220212210232222202210022003301222233200010222
312332322002322000030303223333000233023310233331212333003012112030200010122
200303312...

*(D) Threshold decomposition*: We have decomposed the four-colored image $f(x, y)$ into four binary images $f^i(x, y) = z$ (*Fig. 2A-D*) for a DNA through the threshold decomposition function defined as:

$$f^i(x, y) = 1 \; ; \; z = i : i = 0, 1, 2 \text{ and } 3.$$

$$= 0 \; ; z \neq i$$

Those decomposed binary images for one human X and Y are denoted as $f_{SRY}{}^A, f_{SRY}{}^T, f_{SRY}{}^G$ *and* $f_{SRY}{}^C$ are shown in the following:

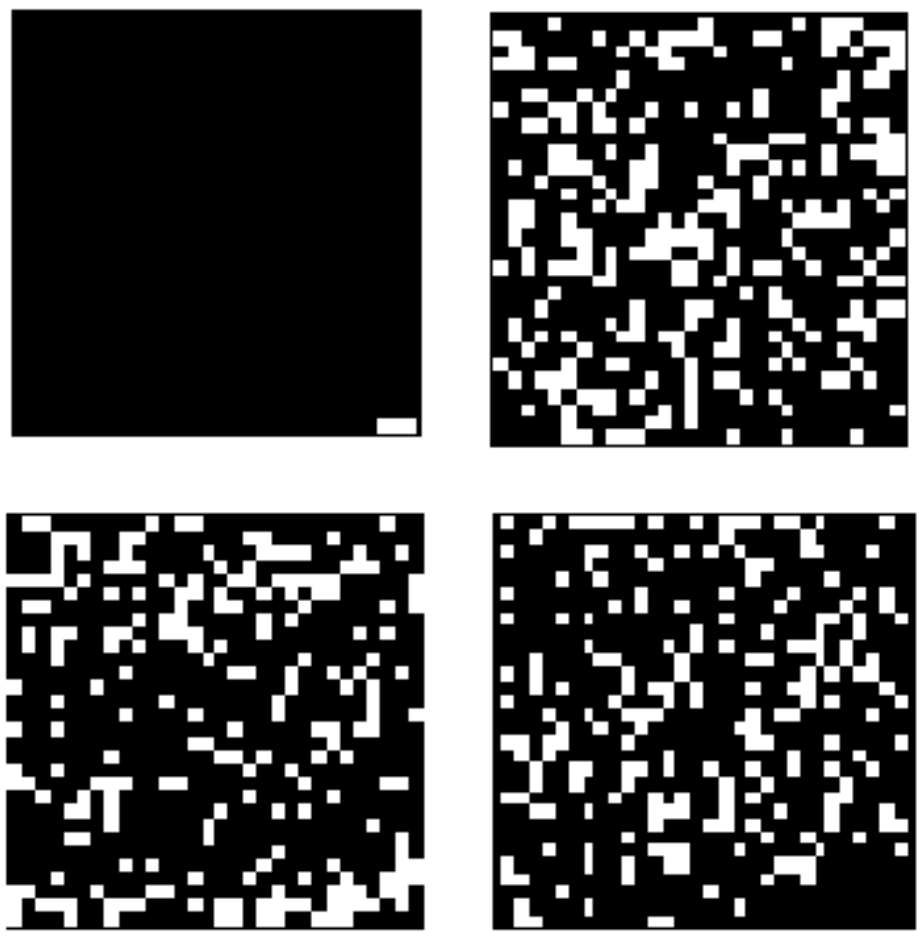

*Fig. 2: Threshold decomposed binary images of OR10AB1P (Black and white denotes complimentary space and one of the ATGC). (A) A (B) T (C) G, and (D) C.*

*(E) Skeleton decomposition*: Morphological skeletons *Fig. 5.6 a-d* for threshold decomposed binary images of SRY shown in (*Fig. 5.6 a-d)* is obtained according to (3).

$SK(X) = [(X \ominus nB)(X \ominus nB)$ O B] …… (3) where $B$ is a structuring element that is symmetric about the origin, and $nB = B \oplus B \oplus B \ldots \oplus B(n\ times)$.

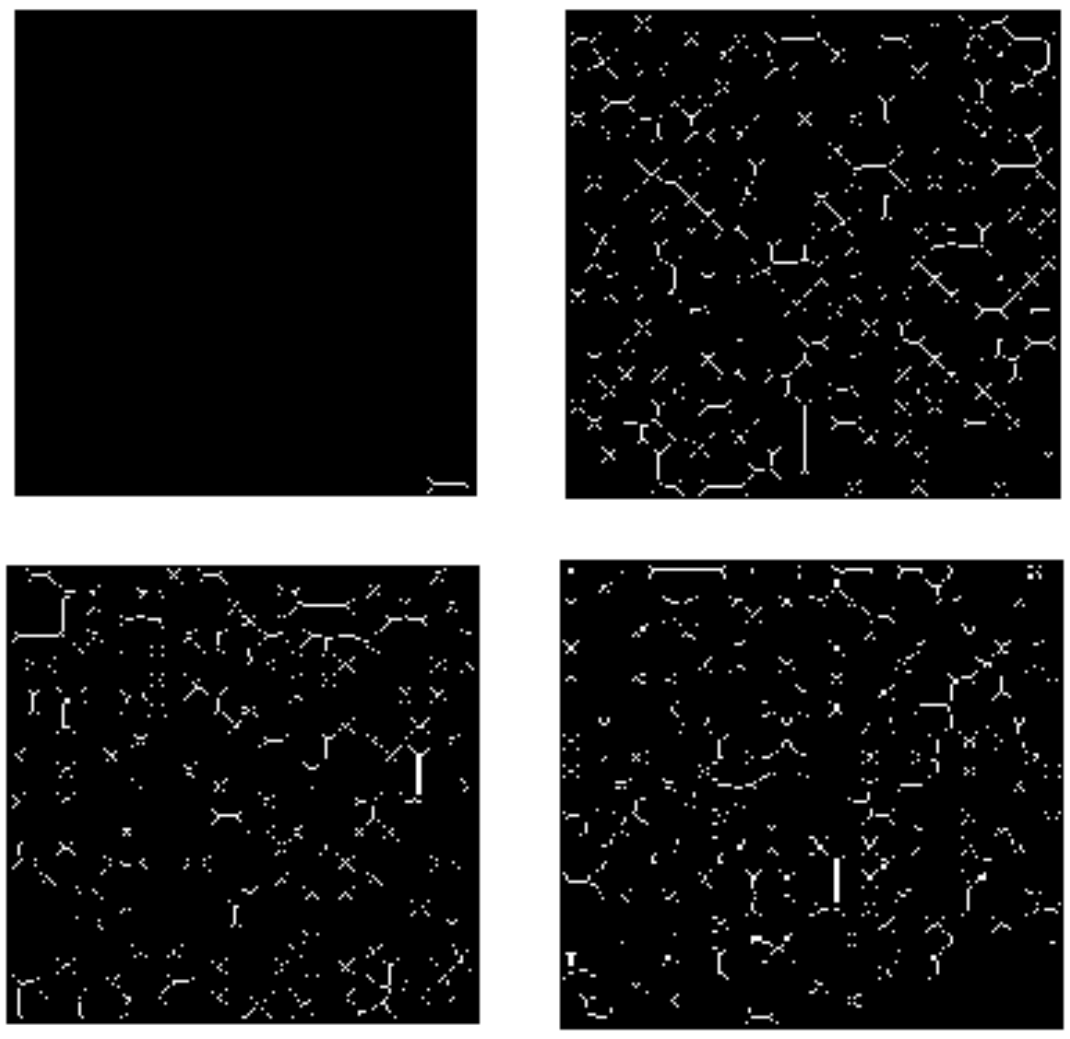

*Fig. 3 Skeletons of* ${f_{SRY}}^{A}, {f_{SRY}}^{T}, {f_{SRY}}^{G}$ *and* ${f_{SRY}}^{C}$

Density and intricacy of the skeleton for those decomposed binary images depend upon the frequency of occurrence of nucleotide chosen as threshold and their spatial distribution. The intricacy of the skeleton is proportional to the heterogeneity in the spatial distribution of the skeleton.

*(F) Protein Plot Representations*:

A number of fundamental properties namely percentage of Accessible Residues (AR), Buried Residues (BR), Alpha Helix (Chou & Fasman) (ALCF), Amino Acid Composition (ACC), Beta Sheet (Chou & Fasman) (BSCF), Beta Turn (Chou & Fasman) (BTCF), Coil (Deleage & Roux) (CDR), Hydrophobicity (Aboderin) (HA), Molecular Weight (MW), and Polarity (P) of the human X and Y chromosomal genes have been considered.

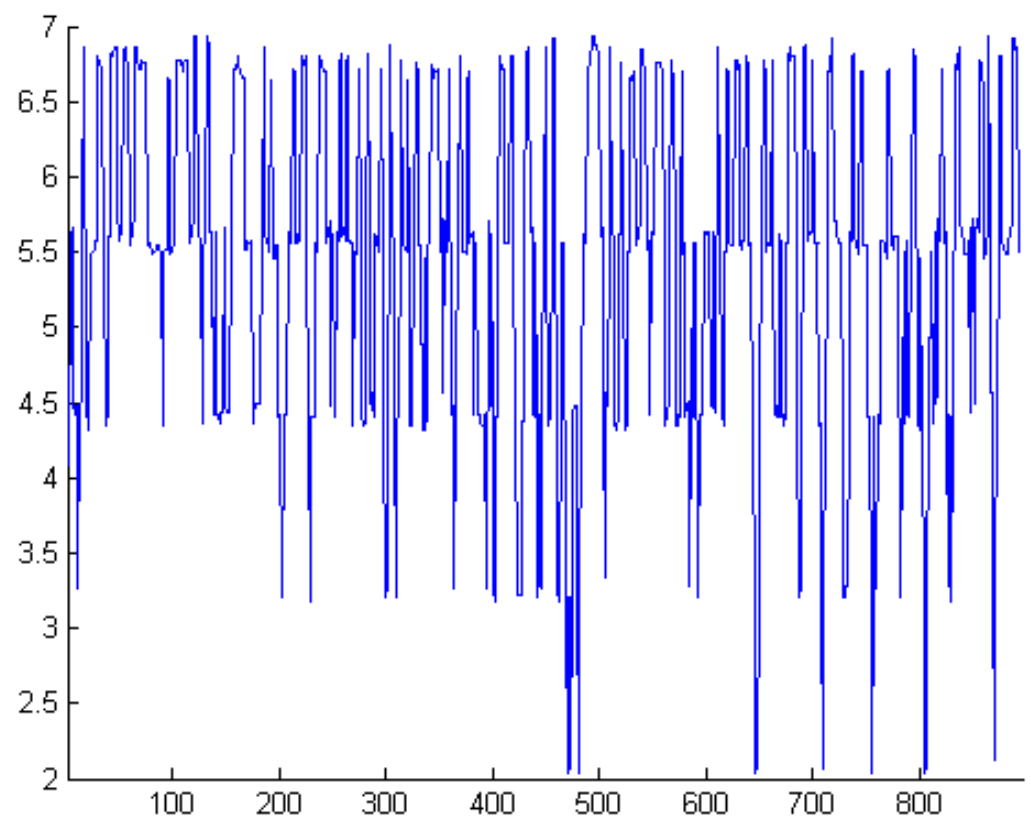

*Fig. 4: Accessible residues (AR-Protein Plot) of SRY*

All protein plots are generated from the gene sequences using Matlab (*bioinformatics toolbox*) (*Fig. 4*). Then box-counting dimension for each of the protein plot have been calculated through BENOIT$^{TM}$.

In the next section let us elaborate the methods applied to DNA string to extract the quantitative details.

## 2. Methods

The quantitative details of X and Y chromosomal genes have been studied in the light of fractal dimension. The very basic of one such fractal dimension method is *Box Counting Dimension* which is illustrated below.

*Box-Counting Method:*

The most practical and commonly used method of calculation of fractal dimension is Box-counting dimension. This is mainly because it is easy to calculate mathematically and because it is easily estimated empirically. We note that the number of line segments of length $\delta$ that are needed to cover a line of length l is $\frac{1}{\delta}$, that the number of squares with side length δ that are needed to cover a square with area A is $\frac{A}{\delta^2}$, and that the number of cubes with side length δ that are needed to cover a cube with volume V is $\frac{V}{\delta^3}$,.

So in general, the box-counting dimension (or just ``box dimension") of a set S subset of $\mathbb{R}^n$ as follows:

For any $\varepsilon > 0$, let $N_\varepsilon(S)$ be the minimum number of n-dimensional cubes of side-length ε needed to cover S. If there is a number D so that

$$N_\varepsilon(S) = 1/\varepsilon^D$$

The D is called the **Box-Counting Dimension** of S.

It is to be Noted that the Box-Counting Dimension is D if and only if there is some positive constant m so that

$$\lim_{\varepsilon\to 0}\frac{N_\varepsilon(S)}{1/\varepsilon^D} = m$$

The above equation gives $D = \lim_{\varepsilon\to 0}\frac{\log m - \log N_\varepsilon(S)}{\log \varepsilon} = -\lim_{\varepsilon\to 0}\frac{\log N_\varepsilon(S)}{\log \varepsilon}$.

Note that the log m term drops out, because it is constant while the denominator becomes infinite as $\varepsilon \to 0$. Also, since $0 < \varepsilon < 1$, log ε is negative, so D is positive.

But in practice, this method computes the number of cells required to entirely cover an object, with grids of cells of varying size. Practically, this is performed by superimposing regular grids over an object and by counting the number of occupied cells. The logarithm of $N(r)$, the number of occupied cells, versus the logarithm of $1/r$, where r is the size of one cell, gives a line whose gradient corresponds to the box dimension [14].

## 2.1 FD of DNA walks of the genes

The DNA walk is defined as a sum of the progression $\sum D_n, n = 1,2,\ldots\ldots,N$ & $D_n \epsilon\{1,2,3,4\}$ which is the cumulative sum on the DNA string representation $\{D_1, D_1 + D_2, \ldots\ldots, \sum_{m=1}^{n-1} D_m, \ldots\ldots, \sum_{m=1}^{N} D_m\}$ [19].

Also we define $a_n \stackrel{\text{def}}{=} \sum_{i=1}^{n} f(A, x_i), g_n \stackrel{\text{def}}{=} \sum_{i=1}^{n} f(G, x_i), c_n \stackrel{\text{def}}{=} \sum_{i=1}^{n} f(C, x_i)$ & $t_n \stackrel{\text{def}}{=} \sum_{i=1}^{n} f(U, x_i)$. It has been resulted by plotting $(P_n, Q_n)$ as we have defined two functions:

$$P_n \stackrel{\text{def}}{=} \sin a_n^2 - \sin g_n^2 \text{ and } Q_n \stackrel{\text{def}}{=} \sin t_n^2 - \sin c_n^2.$$

Here we compute the Fractal dimension of all DNA walk for the *4-adic string* of all sex genes. The plot of the DNA walk for the SRY string is shown in *Fig. 5*.

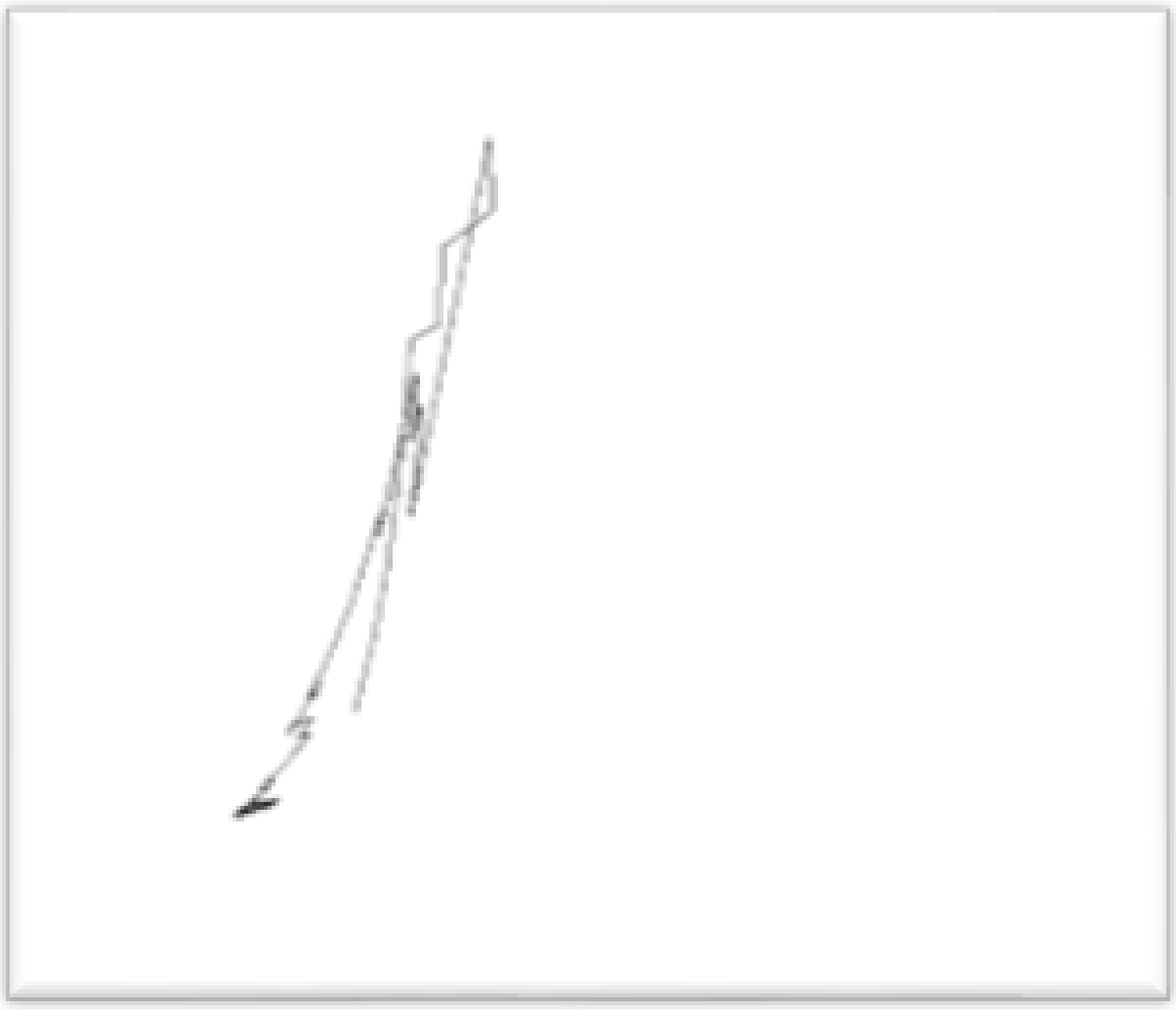

*Fig. 5: DNA Walk $(P_n, Q_n)$ for SRY*

The box-counting dimension for the DNA walk of SRY is 1.94611. In the similar manner we have computed all the box counting dimension of all genes.

## 2.2 Hurst Exponent of the DNA sequences

Hurst exponent is referred as the "index of dependence," and is the relative tendency of a time series either to regress strongly to the mean or to cluster in a direction. It is a measure of long range correlation of one-dimensional time series [19, 20].

Let us consider a string $H = \{h_i\},\ i = 1,2,\dots,n$

$$m_{X,n} = \frac{1}{n}\sum_{i=1}^{n} h_i$$

$$H(i,x) = \sum_{j=1}^{i} \{h_j - m_{x,n}\}$$

$$R(n) = \max H(i,n) - \min H(i,n) \quad 1 \le i \le n$$

$$S(n) = \sqrt{\frac{1}{n}\sum_{i=1}^{n} \left(h_i - m_{x,n}\right)^2}$$

The Hurst exponent H is defined as: $(\frac{n}{2})^H = \frac{R(n)}{S(n)}$, where $n$ is the length of the string. The range for which the Hurst exponent, H indicates negative, positive auto-correlation are $0 < H < 0.5$ and $0.5 < H < 1$ respectively. A value of $H=0.5$ indicates a true random walk, where it is equally likely that a decrease or an increase will follow from any particular value [20].

Here we consider 2-adic strings of DNA for computation of Hurst exponent.

The Hurst exponents of the *2-adic string* of SRY are 0.69106. This is how we have computed Hurst exponent for all the human genes [18].

## 2.3 Succolarity

The degree of percolation of an image (how much a given fluid can flow through this image) can be measured through Succolarity, a fractal parameter [21].

The succolarity of a binary image is defined as

$$\sigma(BS(k), dir) = \frac{\sum_{k=1}^{n} OP(BS(k)) \times PR(BS(k), pc)}{\sum_{k=1}^{n} PR(BS(k), pc)}$$

where '$dir$' denotes direction; $BS(n)$ where n is the number of possible divisions of a binary image in boxes. The occupation percentage (OP) is defined as, for each box size, k, the sum of the multiplications of the $OP(BS(k))$, where k is a number from 1 to n, by the pressure $PR(BS(k), pc)$, where $pc$ is the position on x or y of the centroid of the box on the scale of pressure) applied to the box are calculated. Therefore for any binary decomposed images of $f(x,y)$, the succolarity can be obtained.

Here we compute succolarity of the decomposed images for DNA as shown in the previous section. The succolarity of the four decomposed images $f_{SRY}^{A}, f_{SRY}^{T}, f_{SRY}^{G}$ and $f_{SRY}^{C}$ of SRY are 0.000351, 0.000782, 0.000272 and 0.000267 respectively.

Similarly, we have computed the succolarity of the decomposed images for all sex genes.

**2.4 Statistical Autocorrelations**

It is one of several descriptors, describing how far the values lie from the mean (expected value).

For a given sequence $\{Y_1, Y_2 \ldots Y_N\}$,

$\sigma^2 \stackrel{\text{def}}{=} \frac{1}{N}\sum_{i=1}^{N} Y_i^2 - (\frac{1}{N}\sum_{i=1}^{N} Y_i)^2$ and the variance at distance N-k is given as

$$\sigma^2 \stackrel{\text{def}}{=} \frac{1}{N-k}\sum_{i=1}^{N-k} Y_i^2 - (\frac{1}{N-k}\sum_{i=1}^{N-k} Y_i)^2$$ [**11**].

It is easily computable that the variance for the string *Fig. 1(C)* is 1.29.

**2.5 Mean and SD Ordering of Gene Sequences**

A gene is a string constituting of different permutations of the base pairs A, C, T and G where repetition of a base pair is allowed. We can classify the miRNA sequences based on the ordering of poly-string mean of A, C, T, and G in the string. Given a string *X*, we calculate the mean of poly-strings consisting only of A, C, T and G separately [15, 16].

Mean Nu $= 2(\mathrm{Nu}_1 + \mathrm{Nu}_2 + \mathrm{Nu}_3 + \mathrm{Nu}_4 \ldots + \mathrm{Nu}_m)/\mathrm{m}.(\mathrm{m}+1)$ where $\mathrm{Nu}_i \in \{A, U, C, G\}, i = 1,2, \ldots, m$ and m is the length of the longest poly-string over the string.

According to the non-decreasing order of mean, we have classified all the genes into different classes. The mean order of sequence *1(A)* is AUGC i.e. mean of poly-string of A is less than the same of U and so on.

## 3. Results and Discussions

Let us now elaborate in detail, the result obtained for all sex genes using the above stated methods.

### 3.1 Fractal Dimension of DNA Walk

For all the 92 human sex chromosomal genes along with their homologues in other species, the fractal dimension of the DNA walks lies in the interval (1.896, 1.948) as shown in *Tab.1* and *Tab. 2*.

| Genes | FD of DNA Walk |
|---|---|
| DHRSX (*Bos taurus*) | 1.94577 |
| DHRSX (*Homo sapiens*) | 1.94584 |
| CD99 (*Bos taurus*) | 1.94527 |
| CD99 (*Homo sapiens*) | 1.92993 |
| CD99 (*Pan troglodytes*) | 1.94588 |
| CD9912 (*Mus musculus*) | 1.94584 |
| ZBED1 (*Bos taurus*) | 1.94577 |
| ZBED1 (*Homo sapiens*) | 1.92829 |
| PRKX (*Homo sapiens*) | 1.94595 |
| PRKX (*Pan troglodytes*) | 1.94573 |
| PRKY (*Homo sapiens*) | 1.94595 |
| NLGN4Y (*Homo sapiens*) | 1.90746 |
| NLGN4Y (*Pan troglodytes*) | 1.94615 |
| VCX2 (*Homo sapiens*) | 1.89608 |
| VCX2 (*Pan troglodytes*) | 1.94592 |
| TBL1Y (*Homo sapiens*) | 1.94556 |
| TBL1Y (*Pan troglodytes*) | 1.94579 |
| SC25A6 (*Homo sapiens*) | 1.9298 |
| SC25A6 (*Mus muscullus*) | 1.9458 |
| ASMTL (*Gallus gallus domesticus*) | 1.9457 |
| ASMTL (*Homo sapiens*) | 1.92995 |
| ASMTL (*Canis lupus*) | 1.9458 |
| ASMT (*Homo sapiens)* | 1.9258 |
| ASMT (*Gallus gallus domesticus)* | 1.9459 |
| IL3RA (*Pan troglodytes*) | 1.9458 |
| IL3RA (*Mus muscullus*) | 1.9458 |
| HSFX1 (*Homo sapiens*) | 1.92828 |
| HSFX1 (*Pan troglodytes*) | 1.9454 |
| HSFY2 (*Homo sapiens*) | 1.92982 |
| IL9R (*Homo sapiens*) | 1.92982 |
| IL9R (*Pan troglodytes*) | 1.94576 |
| IL9R (*Macca mullata*) | 1.94571 |
| SYBL1 (*Homo sapiens*) | 1.92956 |
| SYBL1 (*Macca mullata*) | 1.9457 |
| SPRY3 (*Homo sapiens*) | 1.92964 |
| SPRY3 (*Macca mullata*) | 1.93826 |

*Tab.1: Genes and Their FD of DNA Walk*

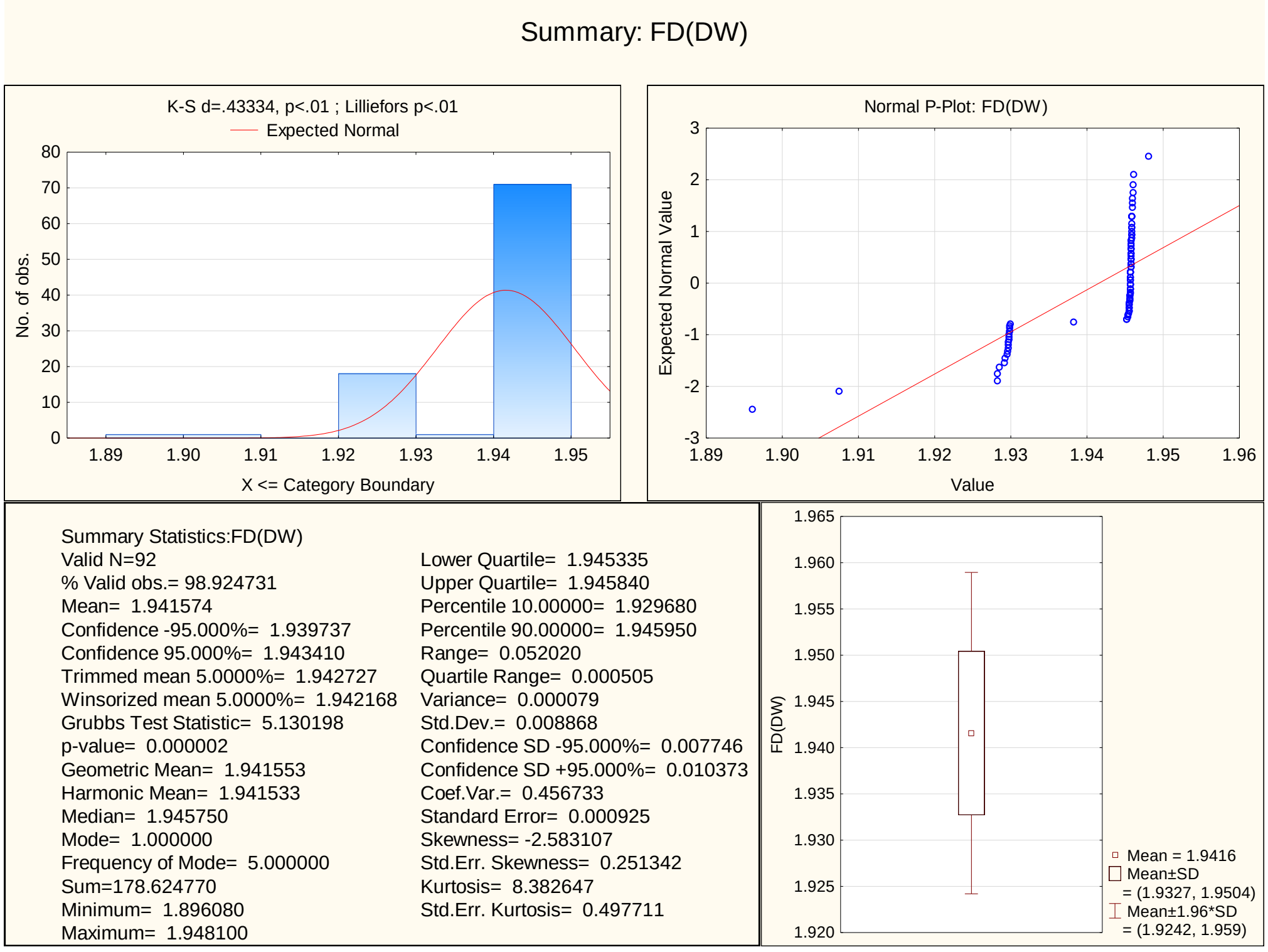

*Tab. 2: Descriptive Statistics of FD of DNA Walk*

From *Tab. 1* and *Tab. 2* it is clear that the box counting dimension is cantered at 1.94. The human sex genes and their corresponding homologues have almost same boxcounting dimensions. Although the ordering and length of the gene sequences are different from each other but they share the same box counting dimension of DNA walk.

### 3.2 Hurst Exponent of 2-adic DNA strings

We have calculated the *Hurst exponent* of all 2-adic strings of DNA for all the human sex genes including their homologues. Hurst exponent lies in the interval (0.5559, 0.98187) and therefore the binary strings of all genes are positively auto-correlated (*Tab. 3*).

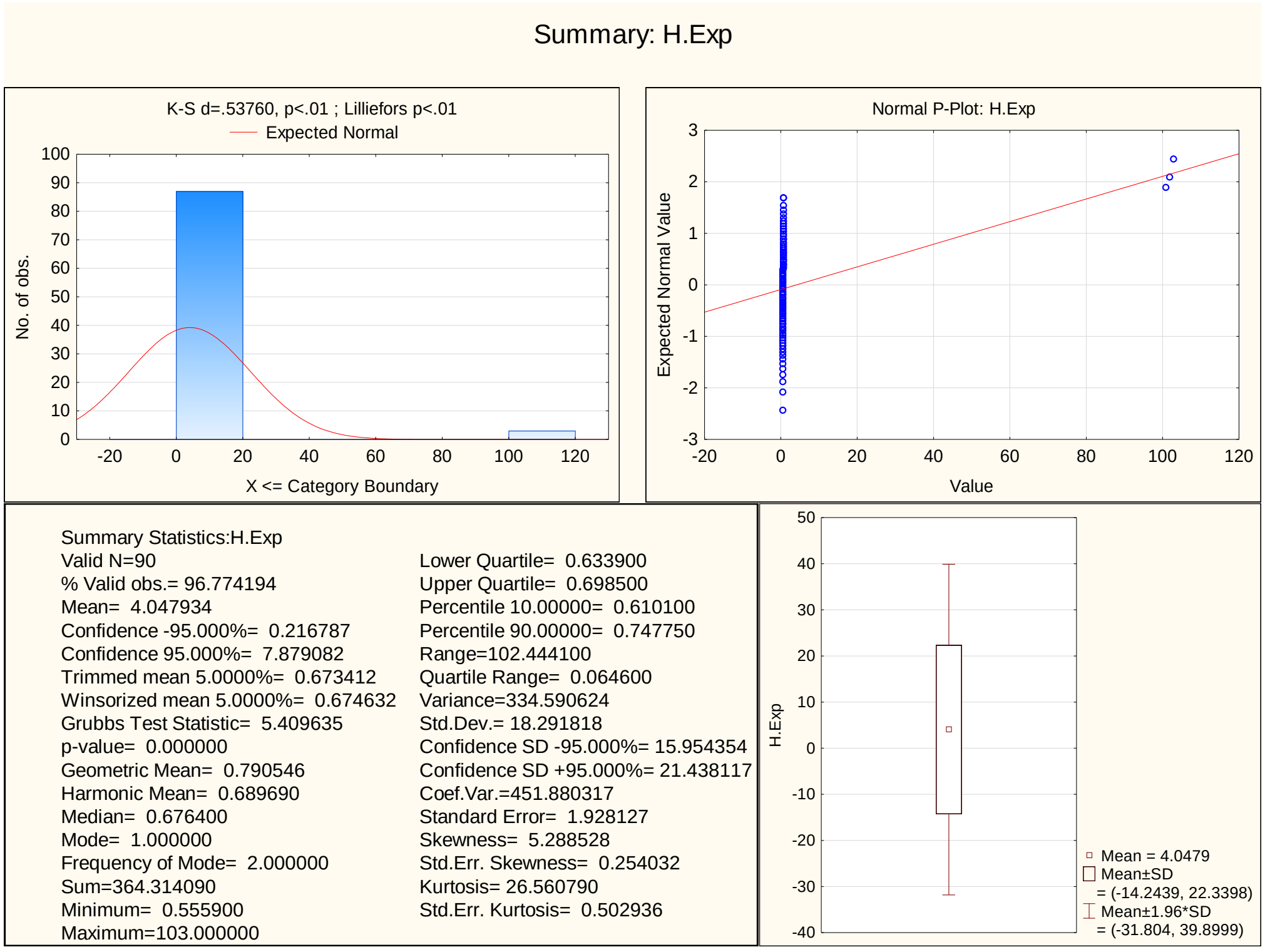

*Tab. 3: Descriptive Statistics of Hurst exponent (2-adic)*

The Hurst exponents for all human sex genes including their corresponding homologue have same as reflected in (*Tab. 3*).

### 3.3 Succolarity Indices

Succolarity measures how much a given fluid can flow through an image, considering as obstacles the set of pixels with a defined color (e.g. white) on 2D images analysis. In other words, it is a measure of continuous density of a 2D pattern.

The succolarity of A for all sex genes lies in the interval (0.000003, 0.2584) and so it is evident that the texture of A for each sex gene is having less density (*Fig. 1*).

The succolarity indices of the genomic textures of T of all the human sex genes with their corresponding homologues are spread over the interval (0.000001, 0.3077) (*Fig. 1*). It is seen that the succolarity i.e. the continuous density of the genomic texture of T is very low as it is seen in case of the genomic texture A . The succolarities for all these 93 genes are centred at 0.03 and there is no much deviation among the succolarity indices.

In case of the genomic texture G, The succolarity indices of all sex genes lie in the interval (0, 1.78) as shown in (*Fig. 6*). The succolarity indices of the genomic texture of C for all genes are computed. It is observed that the indices range from 0 to 0.2459.

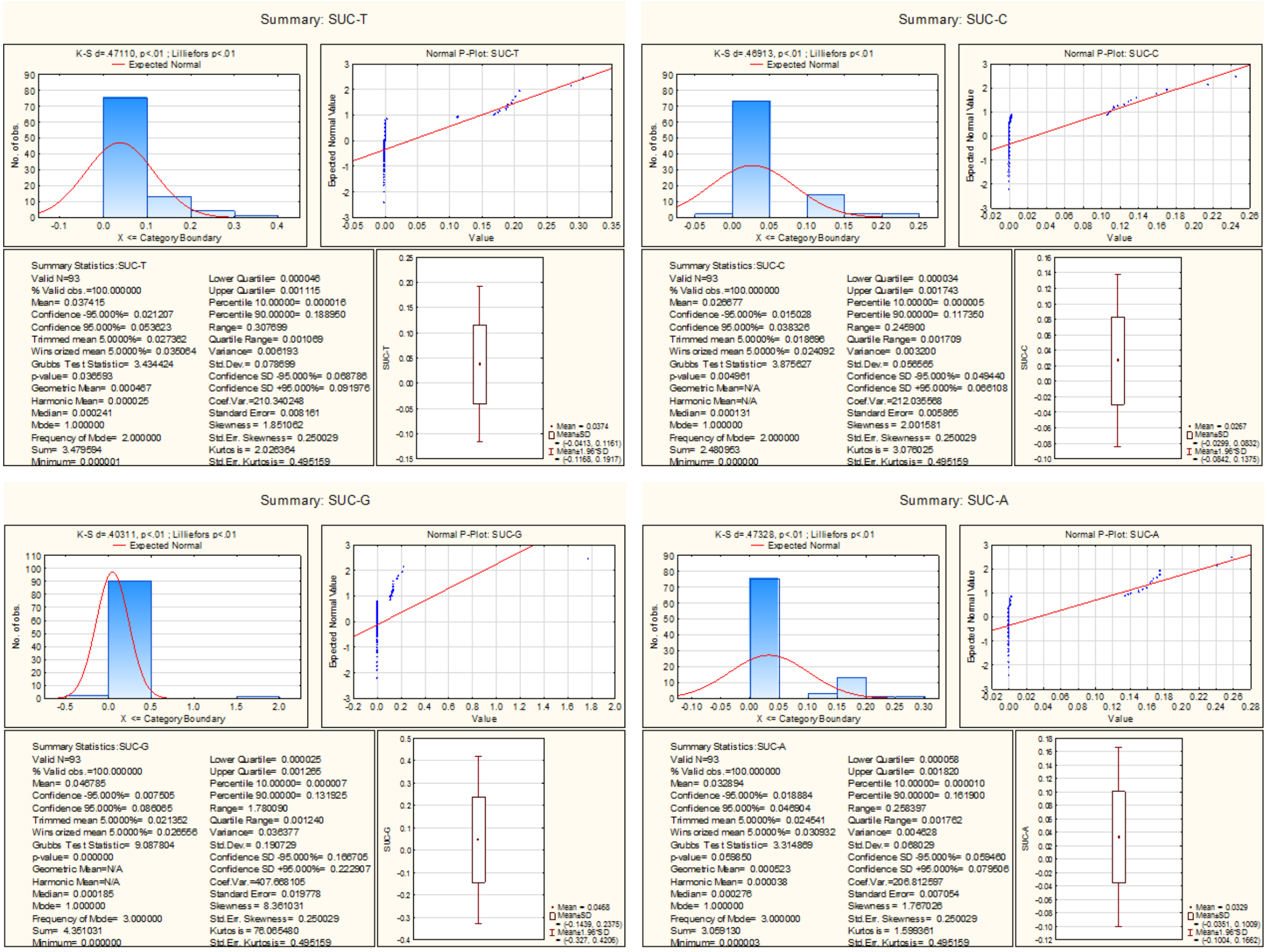

*Fig. 6: Descriptive Statistics of Succolarities of genomic texture A, T, C and G.*

It is seen that the human sex genes and their corresponding homologues share almost same succolarity. Also the succolarity indexes of human sex genes are higher than their corresponding homologues of other species. For an example, the succolarity of A of human sex gene SOY is greater than *Macaca mulatta*'s SOY of monkey and chimpanzee.

The correlation among succolarities of A, T, C and G are illustrated in the *Tab. 4*.

| **Variables** | **SUC-A** | **SUC-T** | **SUC-C** | **SUC-G** |
|---|---|---|---|---|
| **SUC-A** | 1 | 0.994016 | 0.974306 | 0.244261 |
| **SUC-T** | 0.994016 | 1 | 0.957539 | 0.239863 |
| **SUC-C** | 0.974306 | 0.957539 | 1 | 0.251441 |
| **SUC-G** | 0.244261 | 0.239863 | 0.251441 | 1 |

*Tab. 4: Correlation coefficients for Succolarities of A, T, C and G*

The correlation coefficient between the Suc-A and Suc-T is high and in contrast the correlation coefficient between the Suc-G and Suc-C is low.

## 3.4 Statistical Autocorrelations

The statistical autocorrelations ($\sigma$) of the 4-adic representations of all sex genes of human along with their corresponding homologues are being determined. It has been found that the $\sigma$ values lie in the interval (1.06, 1.945).

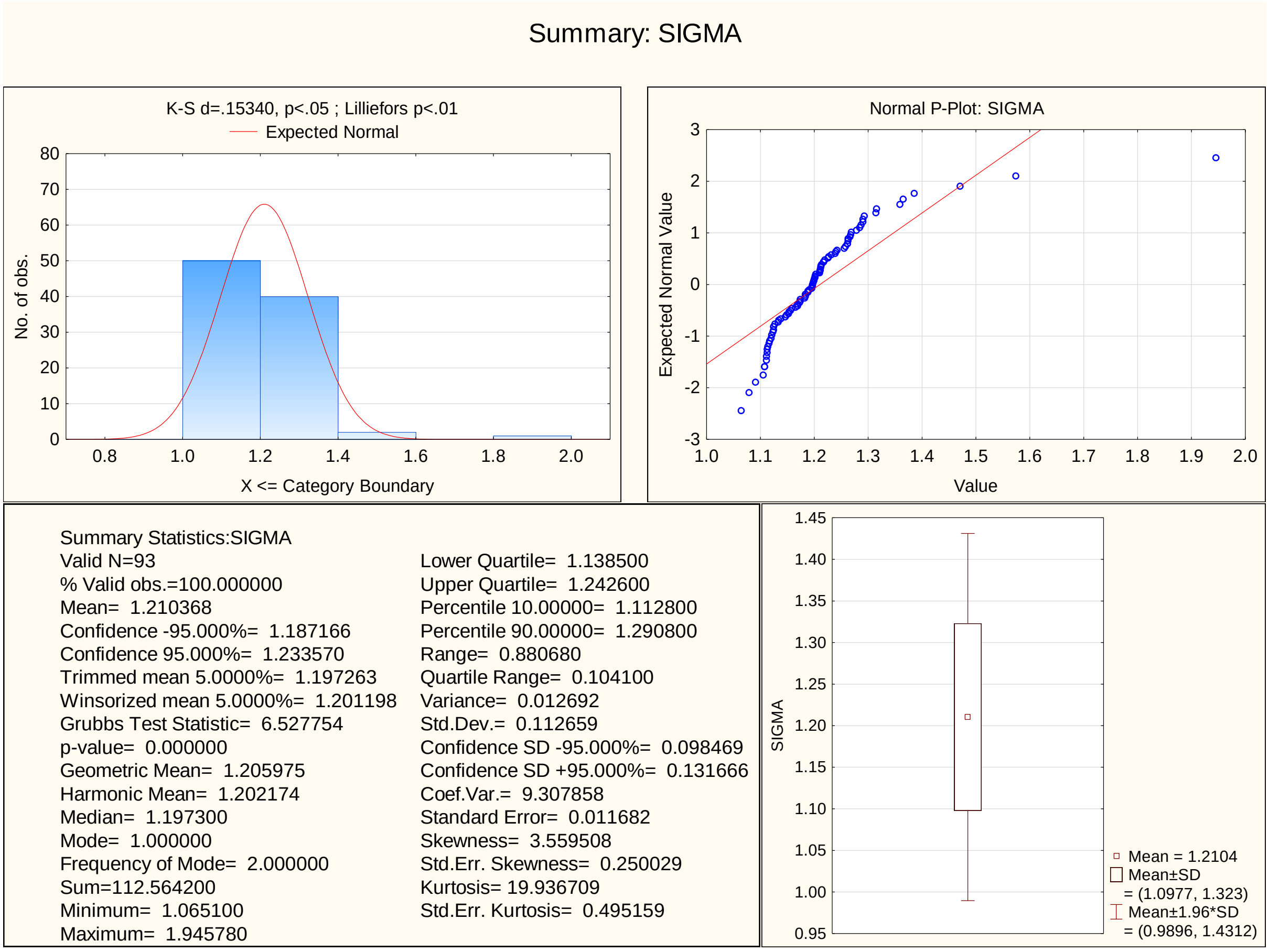

*Tab. 5: Descriptive statistics of statistical autocorrelations*

The sigma values of all sex genes of human and their homologues are normally distributed as shown in the *Tab. 5*.

## 3.5 Fractal Dimensions of Threshold Decomposition Matrices

Here we consider the four different threshold decomposition matrices namely the template of A, T, C and G as we did in the 1.1 (D) for each of the sex-genes (*Fig. 6*). Then we have determined the fractal dimension of the threshold decomposed matrices.

In the *Tab 6*, it is seen that the fractal dimension of the template of A, T, G and C lies in the interval (1.71, 1.81), (1.72, 1.89), (1.70, 1.90) and (1.74, 1.87) respectively.

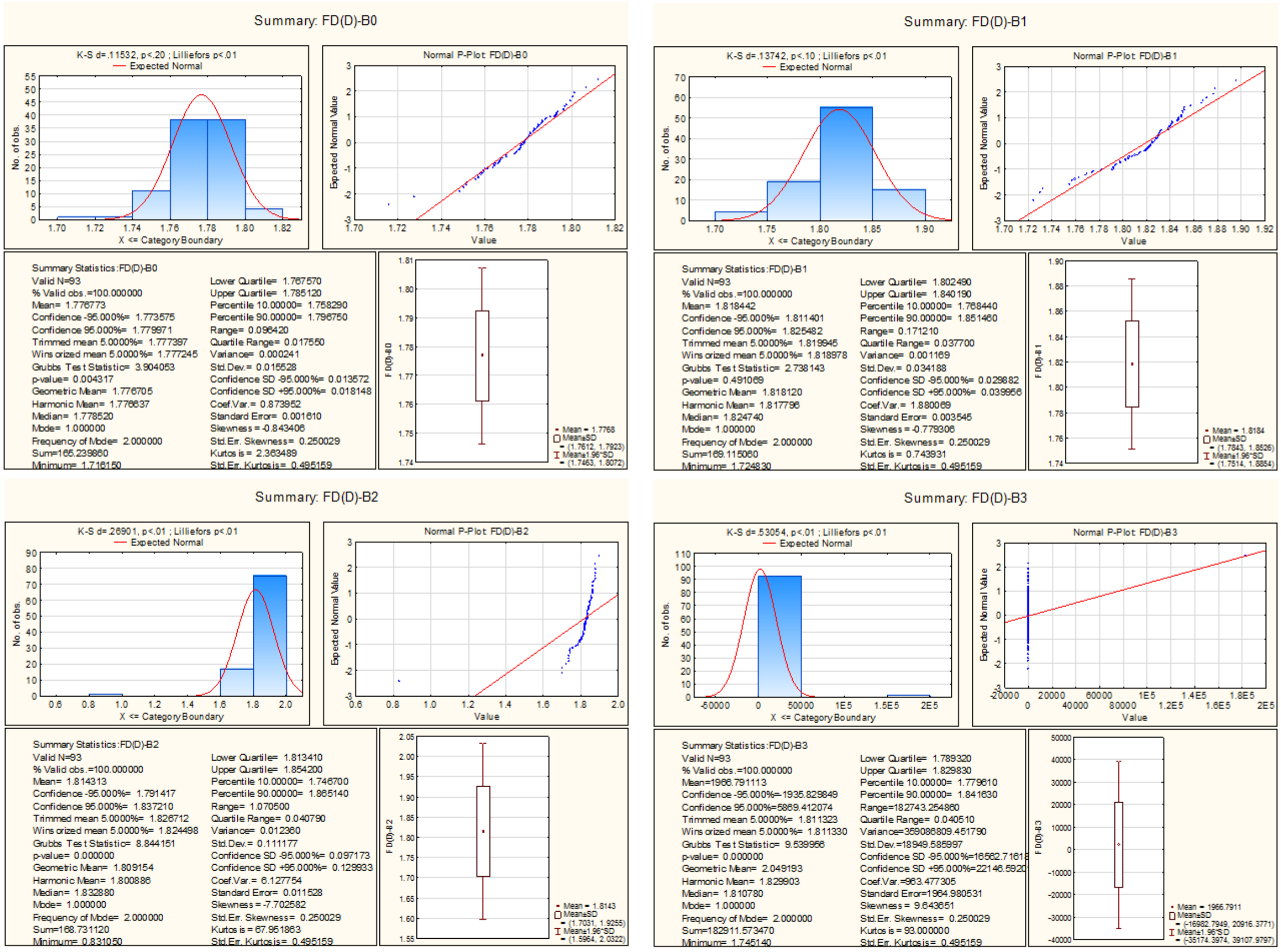

*Tab. 6: Descriptive statistics of fractal dimension of threshold decompositions*

The fractal dimensions of the decomposed template of A and T follow normal distribution whereas fractal dimensions of the other templates do not follow the same.

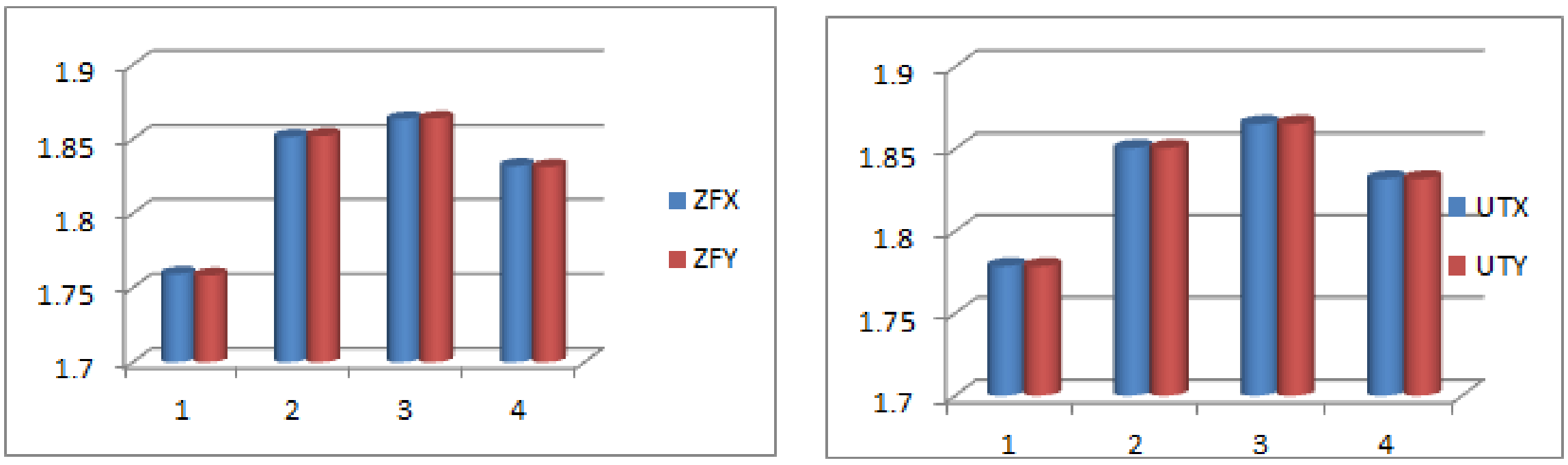

*Fig. 7: Histograms of Fractal dimension of threshold decompositions.*

From the *Fig. 7*, it is seen that the fractal dimensions of the threshold decomposition matrices for A, T, G and C of genes ZFX (from Human X- chromosome) and ZFY (from Human Y-chromosome) are almost same although the genomic template are entirely different in terms of ordering of nucleotides. The aforesaid fact holds good for all the one to one corresponding genes from X and Y chromosomes.

### 3.6 Fractal Dimensions of Skeleton of Threshold Decompositions

In the earlier subsection, the fractal dimensions of threshold decomposition matrices are found. Let us now find out the fractal dimension of the morphological skeleton of all decomposed threshold matrices.

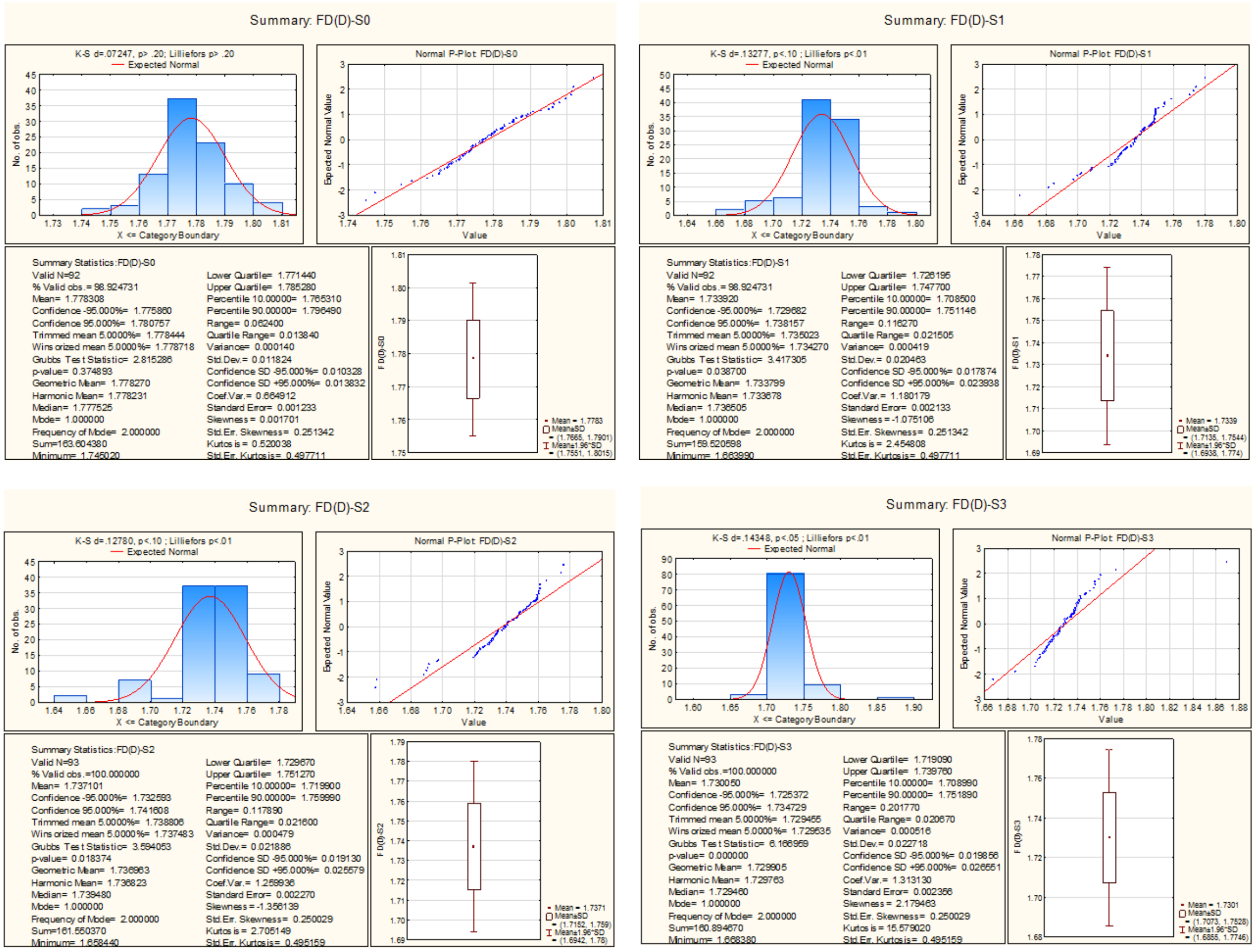

*Tab. 7: Descriptive statistics of fractal dimension of skeleton of threshold decompositions*

In the *Tab 7*, it is determined that the fractal dimensions of the skeletons of template of A, T, G and C lies in the interval (1.74, 1.80), (1.66, 1.78), (1.65, 1.77) and (1.66, 1.87) respectively. The fractal dimensions of the skeleton of decomposed templates of A and T are normally distributed whereas fractal dimensions of the others template do not follow the same as it happened for threshold decomposition matrices.

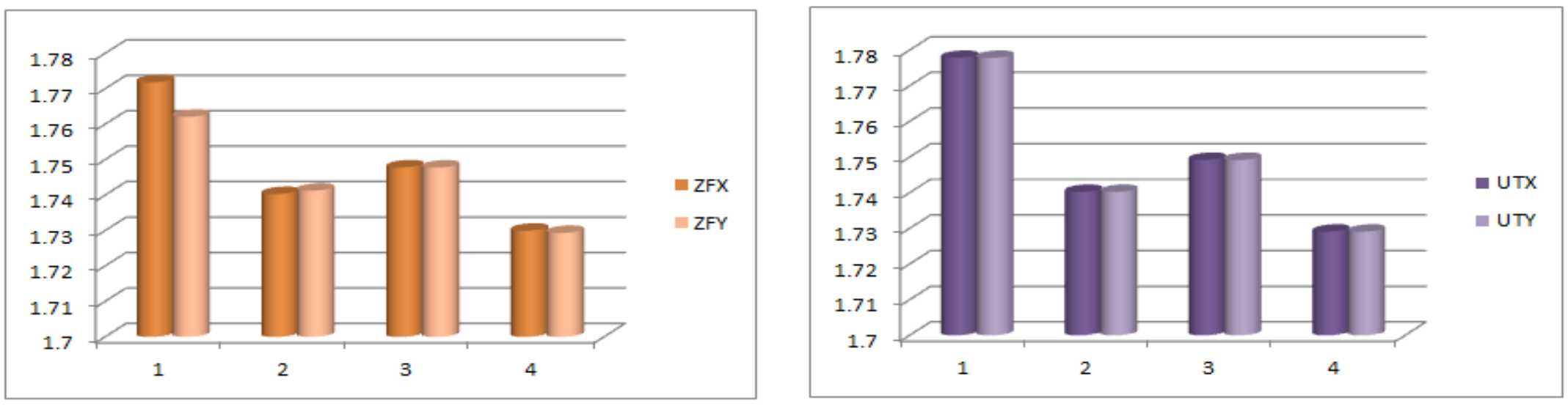

*Fig. 8: Histograms of Fractal dimension of threshold decompositions.*

In cases of ZFX and ZFY genes, the fractal dimensions of the morphological skeletons of the decomposed threshold matrices are almost same. Interestingly, the same is true for all human X and Y chromosomal genes namely as it is evident from the quantitative details (*Supp. Met. 1*).

### 3.7 Fractal Dimension of Protein Plots of Genes

The fractal dimension of the protein plots of Accessible Residues (AR), Buried Residues (BR), Alpha Helix (Chou & Fasman) (ALCF), Amino Acid Composition (ACC), Beta Sheet (Chou & Fasman) (BSCF), Beta Turn (Chou & Fasman) (BTCF), Coil (Deleage & Roux) (CDR), Hydrophobicity (Aboderin) (HA), Molecular Weight (MW), and Polarity (P) of the human X and Y chromosomal genes are lie in the interval (1.81, 1.94), (1.77, 1.94), (1.36, 1.94), (1.79, 1.94), (1.79, 1.94), (1.78, 1.94), (1.84, 1.94), (1.79, 1.94), (1.81, 1.94), (1.81, 1.94), (1.88, 1.94) and (1.81, 1.94) respectively as shown in the descriptive statistics *Fig. 9*.

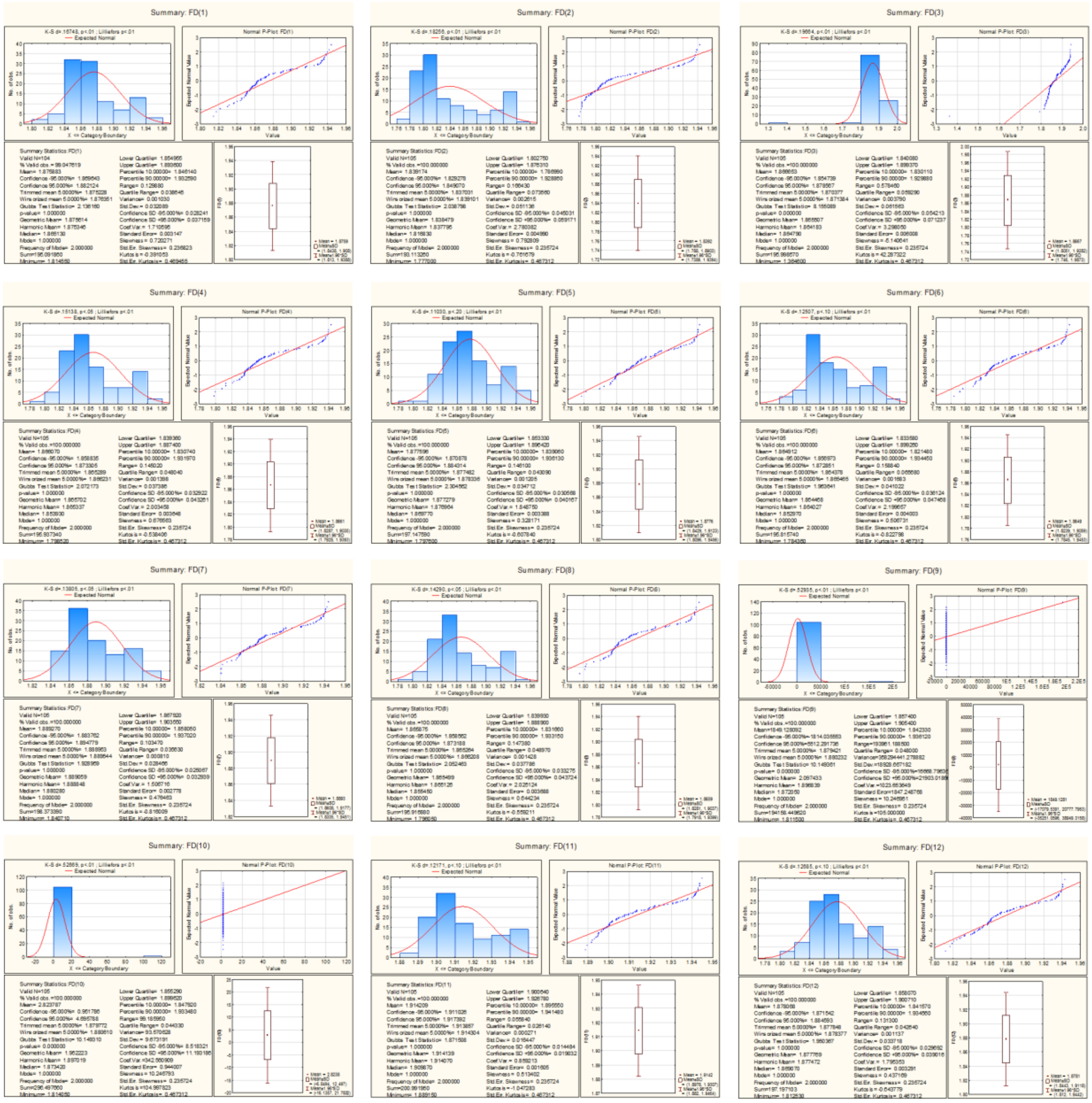

*Fig. 9: Descriptive Statistics of fractal dimension of Protein Plots*

The fractal dimensions of the protein plots of Accessible Residues (AR), Buried Residues (BR), Amino Acid Composition (ACC), Beta Sheet (Chou & Fasman) (BSCF), Beta Turn (Chou & Fasman) (BTCF), Molecular Weight (MW) and Polarity (P) are following normal distribution (*Fig. 9*).

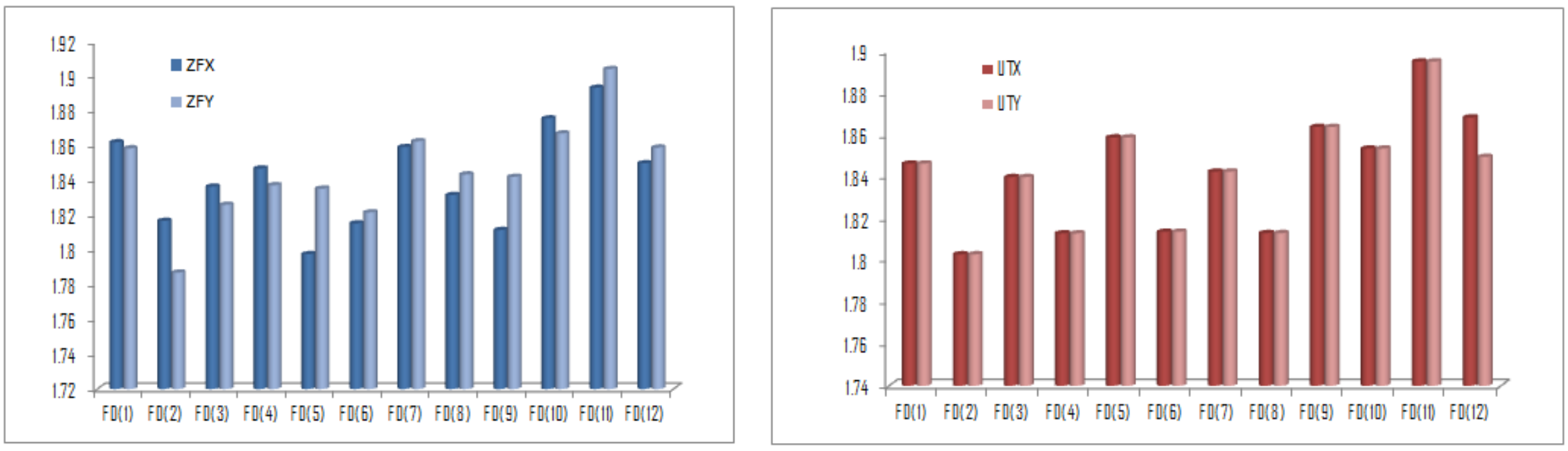

*Fig. 10: Fractal dimension of Protein Plots of ZFX, ZFY and UTX, UTY.*

In cases of the gene pairs (ZFX, ZFY) and (UTX, UTY), the fractal dimensions are agreed almost for all of the protein plots except one or two (*Fig. 10*). In our conviction, these non-agreements make them different as X and Y chromosomal gene. Interestingly, this fact is true for all the X and Y chromosomal homologues of human as evident from the quantitative data (*Supp. Met-2*).

### 3.8 Quantitative Classification of Human Sex Genes

All human sex genes including their homologues are classified based on the order of polystring mean ($P_M^N$) and polystring standard deviation ($P_{SD}^N$). The genes are tabulated in the *Suppl. Met. 3A and Supp. Met. 3B*. The polystring mean ( $P_M^N$) and polystring standard deviation ($P_{SD}^N$) of the human sex gene pair (UTX, UTY) is same as tabulated in the *Suppl. Met. 3A. and Supp. Met. 3B*. In contrary, the gene pair (ZFX, ZFY) has the same polystring mean order but polystring SD order of ZFX is TAGC whereas the same of ZFY is TACG. Even for the homologues of CD99 in three different species namely *Homo sapiens, Pan troglodytes and Bos taurus* have the same polystring mean order namely TGCA but the polystring SD of CD99 of *Pan troglodytes and Bos taurus* same but CD99 for human is different from other two. This fact reveals that the ordering of nucleotides in a gene is most important feature to distinguish them from each other and make them unique. The above phenomena are equally true for other human sex genes and their homologues.

## 4. Conclusion and Future Endeavours

In this paper, a quantitative and deterministic detail is adumbrated through which a given string of nucleotides can be inferred as a human X or Y chromosomal gene without seeking any biological experiment. This would help us in screening any given stretch of nucleotides of specific length as a Human sex-gene homologue. This quantitative detail of genomic imprints of sex genes would enable biologist to understand them in more precise way from the very genomic composition level and these understanding are the next challenge of current Genomics. It is noted that the proposed deterministic model is not only meant for human sex genes or its homologue but also can be treated as a standard prototype for other genes and genomes. In our future endeavours we would like to validate the model through the biological experiment.

**Authors Contributions:** *Sk. S. Hassan* conceptualized the problem and experiments and performed entire research with the rest of the authors of the article.

**Acknowledgements:** The author Sk. S. Hassan is indebted to his colleague ***Dr. Sudhakar Sahoo*** and the former Director ***Professor Swadheenananda Pattanayak*** of Institute of Mathematics & Applications, Bhubaneswar for their kind help and suggestions.

**Conflicts of Interest:** We all authors of the manuscript certify that there is no conflict of interest with any financial organization regarding the material discussed in the manuscript.